\documentclass[aps,prb,twocolumn,amsmath,superscriptaddress]{revtex4}
\usepackage{graphicx}
\usepackage{color}
\usepackage{epstopdf}
\usepackage{xcolor}

\begin{document}

\title{Resonant optical control of the spin of a single Cr atom in a quantum dot}

\author{A. Lafuente-Sampietro}
\affiliation{Universit\'{e} Grenoble Alpes, Institut N\'{e}el, F-38000 Grenoble, France}
\affiliation{CNRS, Institut N\'{e}el, F-38000 Grenoble, France}
\affiliation{Institute of Materials Science, University of Tsukuba, Japan}

\author{H. Utsumi}
\affiliation{Institute of Materials Science, University of Tsukuba, Japan}

\author{H. Boukari}
\affiliation{Universit\'{e} Grenoble Alpes, Institut N\'{e}el, F-38000 Grenoble, France}
\affiliation{CNRS, Institut N\'{e}el, F-38000 Grenoble, France}

\author{S. Kuroda}
\affiliation{Institute of Materials Science, University of Tsukuba, Japan}

\author{L. Besombes}\email{lucien.besombes@grenoble.cnrs.fr}
\affiliation{Universit\'{e} Grenoble Alpes, Institut N\'{e}el, F-38000 Grenoble, France}
\affiliation{CNRS, Institut N\'{e}el, F-38000 Grenoble, France}

\date{\today}

\begin{abstract}

A Cr atom in a semiconductor host carries a localized spin with an intrinsic large spin to strain coupling particularly promising for the development of hybrid spin-mechanical systems and coherent mechanical spin driving. We demonstrate here that the spin of an individual Cr atom inserted in a semiconductor quantum dot can be controlled optically. We first show that a Cr spin can be prepared by resonant optical pumping. Monitoring the time dependence of the intensity of the resonant fluorescence of the quantum dot during this process permits to probe the dynamics of the optical initialization of the Cr spin. Using this initialization and read-out technique we measured a Cr spin relaxation time at T=5 K of about 2 microseconds. We finally demonstrate that, under a resonant single mode laser field, the energy of any spin state of an individual Cr atom can be independently tuned by using the optical Stark effect.

\end{abstract}

\maketitle

Individual spins in semiconductor nano-structures are promising for the development of quantum information technologies \cite{Petta2005,Koenraad2011,Veldhorst2015}. Spins trapped in optically active quantum dots (QDs) have attracted strong interest since their coupling to light enables fast spin control and optical coherent control has been demonstrated for electron \cite{Atature2006,Press2008} and hole \cite{Gerardot2008,DeGreve2011} spins confined in QDs. Thanks to their expected longer coherence time, localized spins on individual dopants in semiconductors are also a promissing media for storing quantum information. Optically active QDs containing individual or pairs of magnetic dopants have been realized both in II-VI \cite{Besombes2004,Goryca2009,LeGall2010,LeGall2011,Besombes2012} and III-V \cite{Kudelski2007,Krebs2013} semiconductors. In these systems, since the confined carriers and magnetic atom spins become strongly mixed, an optical excitation of the QD can affect the spin state of the atom offering possibilities for a control of the localized spin \cite{Govorov2005,Reiter2013}. The variety of magnetic transition elements that can be incorporated in semiconductors gives a large choice of localized electronic spins, nuclear spins and orbital momentums with optical addressability \cite{Besombes2004,Kobak2014,Smolenski2016,Lafuente2016}. This approach opens a diversity of applications in quantum information and quantum sensing.

Among these magnetic atoms, chromium (Cr) is of particular interest \cite{Lafuente2016}. It incorporates in II-VI semiconductors as Cr$^{2+}$ carrying an electronic spin S=2 and an orbital momentum L=2. Moreover, most of Cr isotopes have no nuclear spin. This simplifies the spin level structure and its coherent dynamics \cite{Vallin1974}. With bi-axial strain, the ground state of the Cr is an orbital singlet with spin degeneracy of 5. The orbital momentum of the Cr connects its spin to the local strain through the modification of the crystal field and the spin-orbit coupling. This spin to strain coupling is more than two orders of magnitude larger than for elements without orbital momentum (NV centers in diamond \cite{Tessier2014}, Mn atoms in II-VI semiconductors \cite{Lafuente2015}). Cr is therefore a promising $qubit$ for hybrid spin-mechanical systems in which the motion of a mechanical oscillator would be coherently coupled to the spin state of a single atom \cite{Rabl2010,Tessier2014,Ovar2014}. The development of these hybrid systems will require efficient control of the Cr spin.

In this article, we first show that the spin of a Cr atom inserted in a CdTe/ZnTe QD can be prepared by resonant optical pumping. The resonant photoluminescence (PL) of the QD is used for the read-out of the Cr spin. This spin initialization and read-out technique is used to probe the Cr spin relaxation which remains in the $\mu$s range at low temperature. We finally demonstrate that under a strong resonant laser field, the energy of any spin state of a Cr atom can be tuned by using the optical Stark effect. This ensemble of experiments set the basis required for a full optical control of this single spin system and opens the path toward the development of coherent spectroscopy techniques to probe the strained induced coherent dynamics of the magnetic atom.

\begin{figure}[hbt]
\includegraphics[width=3.2in]{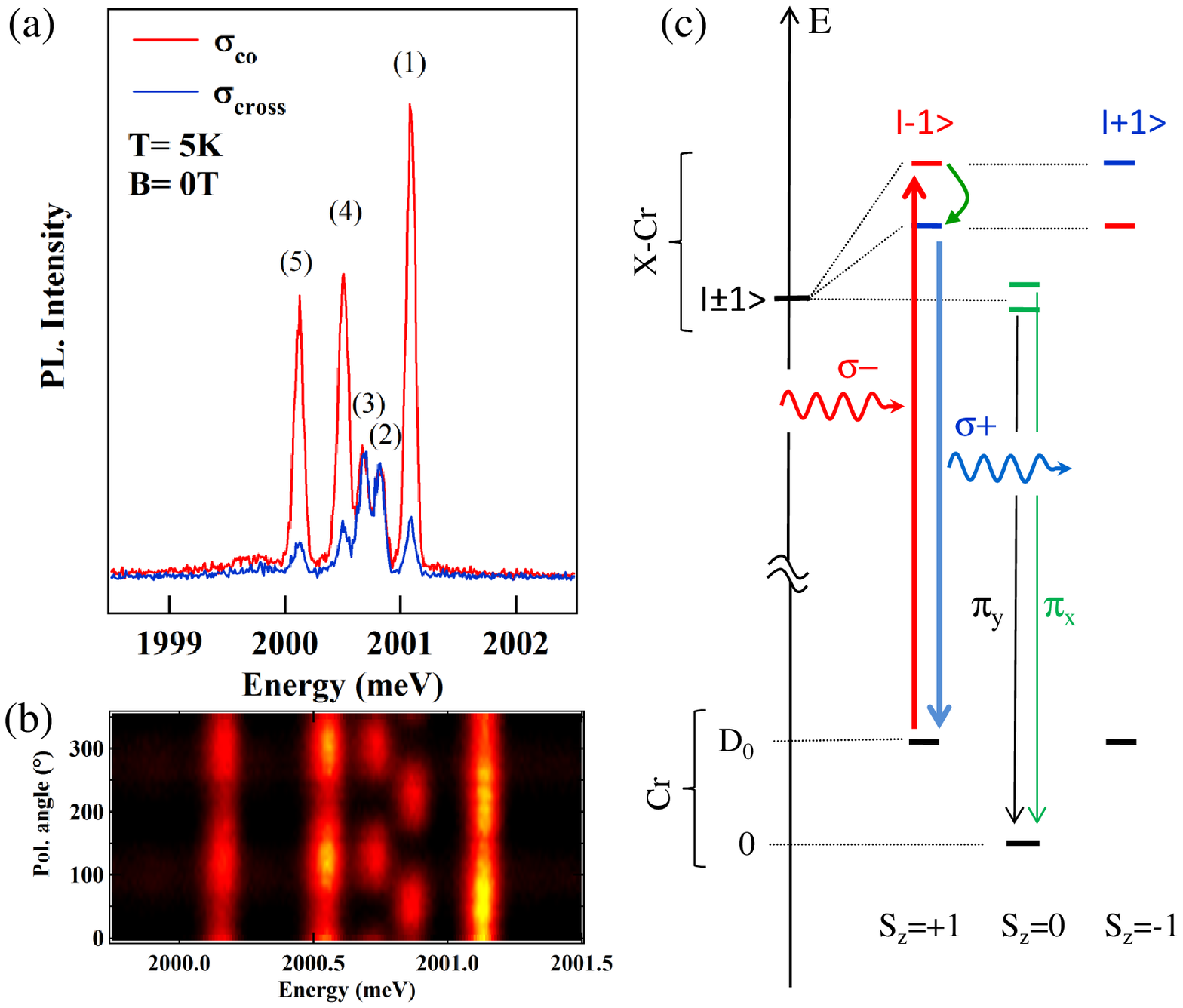}
\caption{(a) PL spectra of the exciton in a Cr-doped QD (X-Cr) for co and cross circularly polarized excitation/detection. (b) Linear polarization intensity map of X-Cr. (c) Energy levels in a Cr-doped QD and configuration of excitation/detection for resonant optical pumping. The Cr states S$_z$=$0$,$\pm$1 are split by the magnetic anisotropy $D_0S_z^2$. In X-Cr, the exchange interaction with the bright exciton ($|\pm1\rangle$) split the states S$_z$=$\pm$1. The higher energy Cr states, S$_z$=$\pm$2, are not displayed.}
\label{Fig1}
\end{figure}

To optically access an individual magnetic atom, Cr are randomly introduced in CdTe/ZnTe self-assembled QDs during their growth \cite{Lafuente2016,Wojnar2011}. The amount of Cr is adjusted to allow the detection of QDs containing 1 or a few Cr atoms. The PL of individual QDs is studied by optical micro-spectroscopy at low temperature (T=5K). The PL is excited with a continuous wave ($cw$) dye laser tuned to an excited state of the QD, dispersed and filtered by a 1 $m$ double spectrometer before being detected by a Si-CCD camera or a fast Si avalanche photodiode. A single-mode dye ring laser can be tuned on resonance with the exciton. In time resolved pumping experiments, trains of resonant light with variable durations and wavelengths are generated from the $cw$ lasers using acousto-optical modulators with a switching time of about 10 $ns$.

The circularly polarized PL spectra of an exciton in a Cr-doped QD (X-Cr) is reported in Fig.~\ref{Fig1}(a). Because of a large magnetic anisotropy of the Cr spin induced by the bi-axial strain in the QDs, $D_0S_z^2$, the Cr spin thermalizes to the states $S_z$=0 and $S_z$=$\pm$1 (Fig.~\ref{Fig1}(c)). The exchange interaction with the spin of the bright exciton $|\pm1\rangle$ further splits the Cr states $S_z$=$\pm$1 (lines (1) and (4)) \cite{Lafuente2016}. Most of the dots also present a low symmetry and the central line associated with $S_z$=0, is split by the electron-hole exchange interaction (lines (2) and (3)) and linearly polarized along two orthogonal directions (Fig.~\ref{Fig1}(b)). An additional line (labeled (5)) appears on the low-energy side of the PL spectra. It arises from a dark exciton $|\pm2\rangle$ which acquires some oscillator strength by a mixing with a bright exciton interacting with the same Cr spin state \cite{Lafuente2016}. This mixing is induced by the electron-hole exchange interaction in a confining potential of symmetry lower than C$_{2v}$ \cite{Zielinski2015}.

To initialize and read-out the Cr spin, we developed a two wavelengths pump-probe experiment. A circularly polarized single mode laser (\emph{resonant pump}) tuned on a X-Cr level is used to pump the Cr spin. Then, a second laser, tuned on an excited state of the QD (\emph{quasi-resonant probe}), injects excitons independently of the Cr spin $S_z$ and drives the Cr to an effective spin temperature where the three ground states $S_z$=0,$\pm$1 are populated \cite{Lafuente2016}. Recording the PL of a X-Cr lines in circular polarization under this periodic sequence of excitation, we monitor the time evolution of the population of a given Cr spin state.

\begin{figure}[hbt]
\includegraphics[width=3.3in]{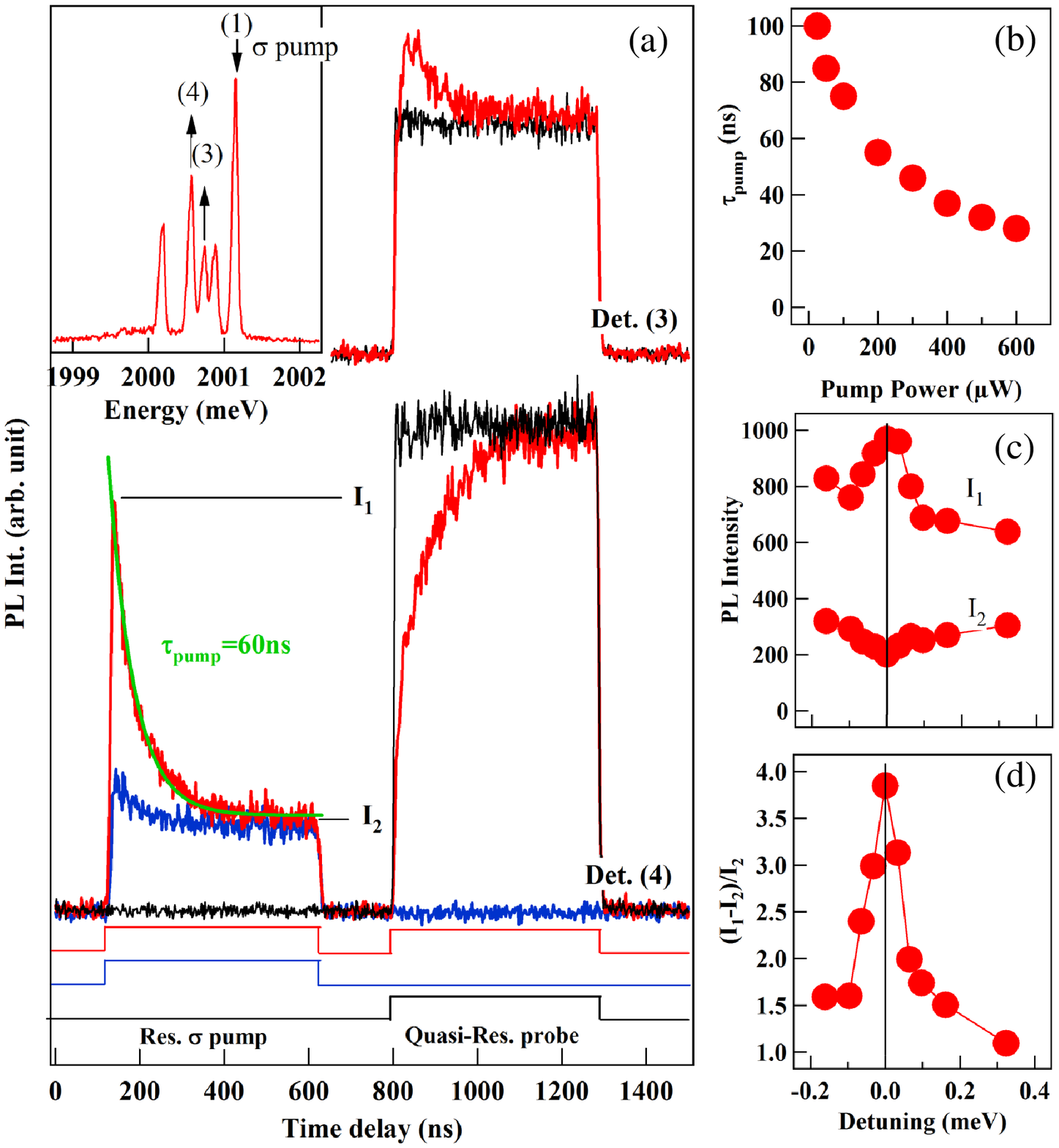}
\caption{(a) PL transients recorded in circular polarization on lines (4) and (3) under the resonant (pump on (1)) and quasi-resonant (probe at E$_{exc.}\approx$2068 meV) optical excitation sequences displayed at the bottom. Inset: PL of X-Cr and configuration of the resonant excitation and detection. (b) Excitation power dependence of the optical pumping time. (c) and (d): Energy detuning dependence of resonant PL intensity (I$_1$, at the beginning and I$_2$, at the end of the pump pulse) and of the corresponding normalized amplitude of pumping transient.}
\label{Fig2}
\end{figure}

The main features of the optical pumping experiment are presented in Fig.~\ref{Fig2}(a). The QD is excited on the high energy state of X-Cr with $\sigma-$ photons. This excitation can only create an exciton in the dot if the Cr spin is $S_z$=+1. An absorption followed by spin-flips of the Cr in the exchange field of the exciton progressively decreases the population of $S_z$=+1. After this pumping sequence, the resonant pump is switched off and followed by the non-resonant probe.

A clear signature of the optical pumping appears on the time evolution of the PL intensity of the low energy line (4). The PL of this line during the probe pulse, recorded in opposite circular polarization with the resonant pump, depends on the population of $S_z$=+1. It strongly differs between the two pump-probe sequences where the resonant pump is on or off. The difference of intensity at the beginning of the probe pulse is a measurement of the efficiency of the pumping. The PL transient during the probe pulse corresponds to a destruction of the non-equilibrium population distribution prepared by the pump. The speed of this spin heating process depends on the intensity of the probe laser. As expected for an increase of the Cr spin temperature, the population of the ground spin state $S_z$=0 also decreases during the probe pulse. This decrease directly appears in the time evolution of the amplitude of the central X-Cr lines during the probe pulse (Det. (3) in Fig.~\ref{Fig2}(a)).

A more direct way to probe the optical pumping speed and efficiency is to monitor the time evolution of the PL during the resonant excitation by the pump pulse. Under resonant excitation on the high energy X-Cr line, an exciton spin-flip with conservation of the Cr spin can produce a PL on the low energy line. This signal depends on the occupation of the Cr state $S_z$=+1 and is used to monitor the time dependence of the spin selective absorption of the QD.

The time evolution of the PL of the low energy line of X-Cr under an excitation on the high energy line is presented in Fig.~\ref{Fig2}(a) for two different pump-probe sequences: probe on and probe off. When the probe laser is on, a large effective Cr spin temperature is established before each pumping pulse. The amplitude of the resonant PL is maximum at the beginning of the pump pulse ($I_1$) and progressively decreases. A decrease of about 80\% is observed with a characteristic decay time in the tens of $ns$ range. As expected for a spin optical pumping process, the characteristic time of the PL transient decreases with the increase of the pump laser intensity (Fig.~\ref{Fig2}(b)). When the probe laser is off, the initial amplitude of the PL transient during the pump pulse is significantly weaker. This decrease is a consequence of the conservation of the out of equilibrium Cr spin distribution during the dark time between two consecutive pumping pulses.

The steady state resonant PL intensity reached at the end of the pump pulse ($I_2$) depends on the optical pumping efficiency which is controlled by the ratio of the spin-flip rate for the Cr spin in the exchange field of the exciton and the relaxation of the Cr spin in the empty dot. However, even with cross-circularly polarized excitation/detection, this steady state PL can also contain a weak contribution from an absorption in the acoustic phonon sideband of the low energy line \cite{Besombes2001}. Fig.~\ref{Fig2}(c) presents the amplitude of the resonant PL detected on the low energy line for a detuning of the pump around the high energy line. A resonance is observed in the initial amplitude $I_1$ of the PL. It reflects the energy and excitation power dependence of the absorption of the QD. A small decrease of the steady state PL $I_2$ is also observed at the resonance. As displayed in Fig.~\ref{Fig2}(d), the corresponding normalized amplitude of the pumping transient, $(I_1-I_2)/I_2$, presents a clear resonant behavior demonstrating the excitation energy dependence of the optical pumping process. The width of the resonance ($\sim 100\mu eV$) is the width of the QD's absorption broadened by the fluctuating environment \cite{Sallen2011}.

\begin{figure}[hbt]
\includegraphics[width=3.2in]{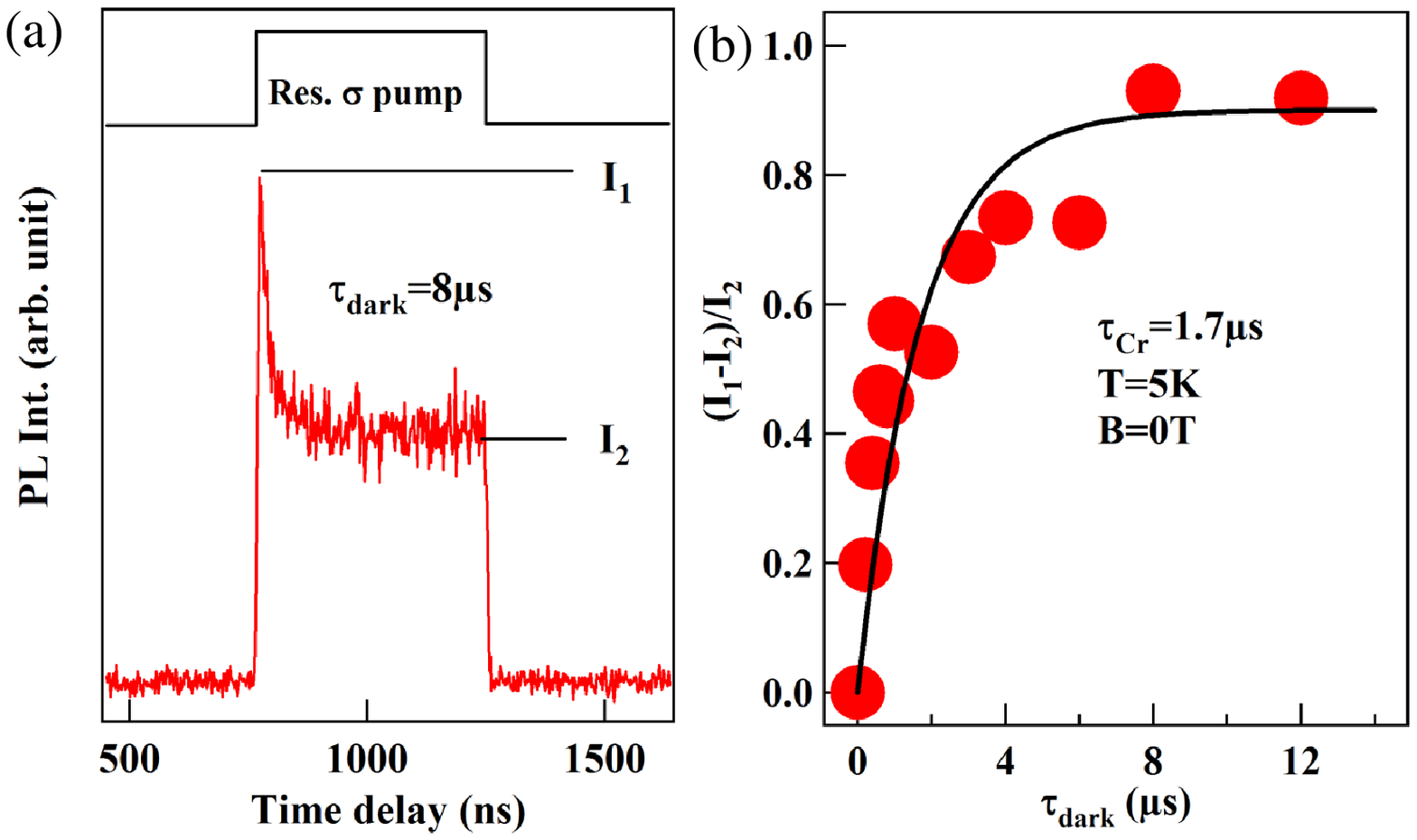}
\caption{(a) Time evolution of the PL intensity of line (4) of X-Cr under resonant excitation on line (1) with a circularly polarized excitation pulse. (b) Evolution of the amplitude of the pumping transient $(I_1-I_2)/I_2$ as a function of the dark time between the excitation pulses. A relaxation time $\tau_{Cr}\approx1.7\mu s$ is measured.}
\label{Fig3}
\end{figure}

With this resonant optical pumping technique to prepare and read-out the Cr spin, we performed pump-probe experiments to observe its relaxation time in the absence of carriers (Fig.~\ref{Fig3}). A non-equilibrium distribution of the Cr spin population is prepared with a circularly polarized resonant pump pulse on the high energy X-Cr line. The pump laser is then switched off, and switched on again after a dark time $\tau_{dark}$. The amplitude of the pumping transient observed on the resonant PL of the low energy line depends on the Cr spin relaxation during $\tau_{dark}$. As presented in Fig.~\ref{Fig3}(b), the amplitude of the transient is fully restored after a dark time of about 10 $\mu$s showing that after this delay the Cr spin is in equilibrium with the lattice temperature (T=5K). Let us note, however, that the initial amplitude of the pumping transient in this case is weaker than the one observed after a non-resonant probe pulse (Fig.~\ref{Fig2}(a)) which drives the Cr spin to an effective temperature much larger than the lattice temperature.

From the time delay dependence of the amplitude of the transient, we deduce a Cr relaxation time $\tau_{Cr}\approx1.7\mu$s at B=0T and T=5K. For such neutral QD and in the absence of optical injection of carriers, this spin relaxation is likely to be controlled by the spin-lattice interaction. Despite the large spin-phonon coupling expected for magnetic atoms with an orbital momentum, the Cr spin relaxation time remains in the $\mu s$ range. This spin memory is long enough for a practical use of Cr in an hybrid spin nano-mechanical system and could even be improved in different QDs structures with weaker biaxial strain \cite{Besombes2014}, lower magnetic anisotropy splitting and consequently less coupling with acoustic phonons \cite{Cao2011}.

These measurements reveal a significantly different Cr spin-flip times under optical excitation and in the dark. The fast Cr spin-flip under optical excitation can be due to the interaction with carriers (exchange induced Cr spin flips) but can also be induced by the interaction with non-equilibrium acoustic phonons created during the energy relaxation of the injected carriers. Both mechanisms probably contribute to the Cr spin heating.

\begin{figure}[hbt]
\includegraphics[width=3.3in]{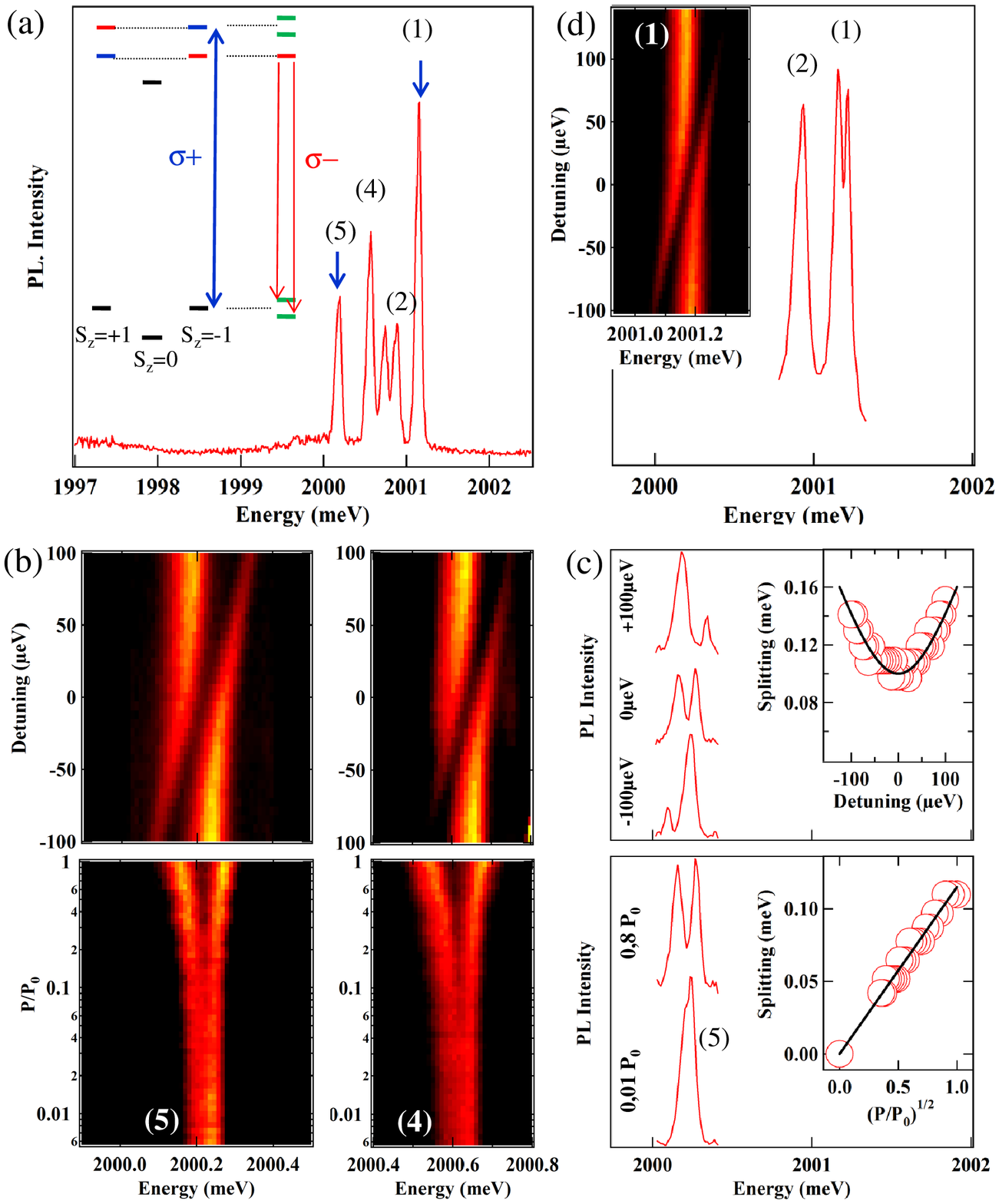}
\caption{(a) PL of X-Cr and configuration of excitation in the resonant optical control experiments. The inset illustrate the laser induced splittings in the ground and excited states for a $\sigma+$ excitation on S$_z$=+1. (b) PL intensity maps of lines (5) and (4) for an excitation on (1) as a function of the detuning (top) and of the excitation intensity (bottom). The PL is produced by a non-resonant laser. The corresponding emission line-shapes are presented in (c) for line (5). The insets in (c) show the splitting of the PL doublet as a function of the excitation intensity (bottom) and laser detuning (top). The fit is obtained with $\hbar\Omega_r$= 100 $\mu eV$. (d) PL of line (1) and (2) for a laser on resonance with the dark exciton state (5). Inset: PL intensity map of line (1) as a function of the laser detuning.}
\label{Fig4}
\end{figure}

The resonant optical excitation on a X-Cr line can also be used to tune the energy of any Cr spin state through the optical Stark effect \cite{LeGall2011}. Such energy shift could be exploited to control the coherent dynamics of the magnetic atom \cite{Jamet2013,Reiter2013}. This optical control technique is presented in Fig.~\ref{Fig4}. When a high intensity single mode laser is tuned to the high energy line of X-Cr in $\sigma+$ polarization, a splitting is observed in $\sigma-$ polarization in the PL of the two low energy lines produced by a second non-resonant laser.

At high excitation intensity, a strong coupling with the resonant laser field mixes the states with a Cr spin component $S_z$=+1 in the presence (X-Cr) or absence (Cr alone) of the exciton. In the ground state of the QD (Cr alone) two hybrid matter-field states are created (inset of Fig.~\ref{Fig4}(a)). Their splitting, $\hbar\Omega_r^{\prime}=\hbar\sqrt{\Omega_r^2+\delta^2}$, depends on the energy detuning of the laser $\hbar\delta$ and on its intensity through the Rabi energy $\hbar\Omega_r$ \cite{Boyle2009}. It can be observed in the PL of all the X-Cr levels associated with $S_z$=+1. The splitting measured on line (5) for a resonant excitation on line (1) is plotted as a function of the square root of the resonant laser intensity in Fig.~\ref{Fig4}(c) showing that, as expected for a two level system, it linearly depends on the laser field strength. The Rabi splitting can reach 150 $\mu eV$ at high excitation power. As the pump laser is detuned, the optically active transitions asymptotically approaches the original excitonic transitions where the remaining offset is the optical Stark shift.

A resonant laser permits to address any spin state of the Cr and selectively shift its energy. For instance, as presented in Fig.~\ref{Fig4}(d), an excitation on the dark exciton state (5) can be used to tune the energy of $S_z$=+1 without affecting $S_z$=0. Such energy tuning is particularly interesting for the control of the Cr spin states $S_z$=$\pm$1 which can be mixed by anisotropic in-plane strain \cite{Lafuente2016}. A relative shift of their energy by a resonant optical excitation would affect their coupling and consequently the Cr spin coherent dynamics. These strong coupling experiments pave the way towards the development of coherent spectroscopy techniques to probe variations of the strain induced mixing of the Cr spin states and makes Cr a promising nano-scale strain sensor.

To conclude, we demonstrated that a resonant optical excitation can be used to control the spin of a Cr atom inserted in a QD. The Cr spin can be initialized by optical pumping, readout through the QD resonant PL and its energy tuned by optical Stark effect. Despite the large spin-lattice coupling expected for a Cr, its spin relaxation time remains in the $\mu$s range. These experiments mark an important step toward the development of optical coherent spectroscopy techniques, such as coherent population trapping, for a precision sensing of applied strain on a Cr inserted in an hybrid spin-mechanical system \cite{Pigeau2015,Barfuss2015}.

\begin{acknowledgements}

The authors acknowledge financial support from the Labex LANEF for the Grenoble-Tsukuba collaboration. The study in Tsukuba has partially been supported by Grant-in-Aid for Scientific Research on Innovative Areas and Exploratory Research "Science of Hybrid Quantum System". The work in Grenoble was realized in the framework of the Commissariat \`{a}  l'Energie Atomique et aux Energies Alternatives (Institut Nanosciences et Cryog\'{e}nie) / Centre National de la Recherche Scientifique (Institut N\'{e}el) joint research team NanoPhysique et Semi-Conducteurs.

\end{acknowledgements}

\end{document}